\documentclass[11pt]{article}
\usepackage{moriond,epsfig}
\usepackage{amssymb}

\def\be{\begin{equation}}
\def\ee{\end{equation}}
\def\bea{\begin{eqnarray}}
\def\eea{\end{eqnarray}}

\def\ifm#1{\relax\ifmmode#1\else$#1$\fi}
\def\DAF{DA\char8NE}  \def\epm{\ifm{e^+e^-}}

\def\ff{$\phi$--factory}

  \def\x{\ifm{\times}}

\def\pt#1,#2,{\ifm{#1\x10^{#2}}}
\newcommand{\kl}{\mbox{$K_L$}}
\newcommand{\ks}{\mbox{$K_S$}}

\newcommand{\Pphi}{\ensuremath{\phi}}

\newcommand{\eV}{{e\kern-.07em V}}
\newcommand{\kloe}{K{\kern-.07em LOE} }
\newcommand{\dafne}{D{\kern-.07em A\ensuremath{\Phi}NE }}

\newcommand{\MeV}{{\rm \,M\eV}}
\newcommand{\GeV}{{\rm G\eV}}
\newcommand{\ps}{{\rm \,ps}}

\newcommand{\mm}{{\rm \,mm}}
\newcommand{\cm}{{\rm \,cm}}
\newcommand{\m}{{\rm \,m}}

\newcommand{\um}{\ensuremath{\mathrm{\mu m}}}
\newcommand{\T}{{\rm \,T}}

\newcommand{\Lfb}{\ensuremath{\rm \, fb^{-1}}}

\newcommand{\DKSpippim}{\ensuremath{K_S\rightarrow\pi^+\pi^-}}

\newcommand{\DKSee}{\ensuremath{K_S\rightarrow e^+e^-}}
\newcommand{\DKLee}{\ensuremath{K_L\rightarrow e^+e^-}}

\newcommand{\DKLmm}{\ensuremath{K_L\rightarrow \mu^+\mu^-}}

\newcommand{\Dphipippimpio}{\ensuremath{\phi\rightarrow\pi^+\pi^-\pi^0}}

\newcommand{\BR}[1]{\ensuremath{\mathrm{BR}(#1)}}

\newcommand{\kcr}{\ensuremath{K_\mathrm{crash}}}

\newcommand{\koko}{\ensuremath{K^0\bar{K}^0}}

\begin{document}
\vspace*{4cm}
\title{Search for the decay \DKSee}

\author{ F. Archilli for \kloe\ collaboration\footnotemark[1] }

\address{Dipartimento di Fisica dell'Universit\`a di Roma Tor Vergata \& sezione INFN Roma Tor Vergata,\\Rome, Italy}

\footnotetext[1]{
F.~Ambrosino,
A.~Antonelli,
M.~Antonelli,
F.~Archilli,
P.~Beltrame,
G.~Bencivenni,
S.~Bertolucci,
C.~Bini,
C.~Bloise,
S.~Bocchetta,
F.~Bossi,
P.~Branchini,
P.~Campana,
G.~Capon,
T.~Capussela,
F.~Ceradini,
F.~Cesario,
P.~Ciambrone,
F.~Crucianelli,
E.~De~Lucia,
A.~De~Santis,
P.~De~Simone,
G.~De~Zorzi,
A.~Denig,
A.~Di~Domenico,
C.~Di~Donato,
B.~Di~Micco,
M.~Dreucci,
G.~Felici,
M.~L.~Ferrer,
S.~Fiore,
P.~Franzini,
C.~Gatti,
P.~Gauzzi,
S.~Giovannella,
E.~Graziani,
W.~Kluge,
V.~Kulikov,
G.~Lanfranchi,
J.~Lee-Franzini,
D.~Leone,
M.~Martini,
P.~Massarotti,
S.~Meola,
S.~Miscetti,
M.~Moulson,
S.~M\"uller,
F.~Murtas,
M.~Napolitano,
F.~Nguyen,
M.~Palutan,
E.~Pasqualucci,
A.~Passeri,
V.~Patera,
F.~Perfetto,
P.~Santangelo,
B.~Sciascia,
A.~Sciubba,
A.~Sibidanov,
T.~Spadaro,
M.~Testa,
L.~Tortora,
P.~Valente,
G.~Venanzoni,
R.Versaci}

\maketitle\abstracts{
We present results of a direct search for the decay \DKSee\ with the 
\kloe\ detector, obtained with a sample of $\epm \to \Pphi \to \ks\kl$ 
events produced at \DAF, the Frascati \ff, for an integrated luminosity
of 1.9~$\Lfb$. The Standard Model prediction for this decay is 
$\BR\DKSee = 2\times 10^{-14}$. The search has been performed by tagging 
the \ks\ decays with simultaneous detection of a \kl\ interaction in the 
calorimeter. Background rejection has been optimized by using both kinematic 
cuts and particle identification. At the end of the analysis chain we find 
$\BR\DKSee < 9.3\times10^{-9}$ at 90\%~CL, which improves by a factor of $\sim 15$ 
on the previous best result, obtained by CPLEAR experiment.}
\section{Introduction}
\begin{figure}[h!]
  \begin{center}
    \epsfig{figure=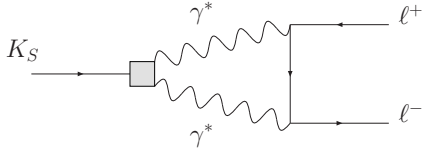,height=0.8in}
    \caption{Long distance contribution to $\ks \rightarrow \ell^+ \ell^-$ process, 
             mediated by two-photon rescattering.}
    \label{fig:ggresc}
  \end{center}
\end{figure}

The decay \DKSee, like the decay  \DKLee\ or \DKLmm, is a flavour-changing 
neutral-current process, suppressed in the Standard Model and dominated by 
the two-photon intermediate state~\cite{kseetheory}.
For both \ks\ and\kl, the \epm\ channel is much more suppressed than the 
$\mu^+\mu^-$ one (by a factor of $\sim 250$) because of the $e-\mu$ mass difference. 
The diagram corresponding to the process $\ks \rightarrow \gamma^* \gamma^* \rightarrow \ell^+ \ell^-$ 
is shown in 
Fig.~\ref{fig:ggresc}.
Using Chiral Perturbation Theory  ($\chi$pT) to order $\mathcal{O}(p^4)$, 
the Standard Model prediction BR$(\DKSee)$
is evaluated to be  $\sim 2 \times 10 ^{-14}$.
A value significantly higher than expected would point to new physics.
The best experimental limit for $BR(\DKSee)$ has been measured
by CPLEAR~\cite{cplearksee}, and it is equal to $1.4\times 10^{-7}$, at $90\%$~CL.
Here we present a new measurement of this channel, which improves on the previous result
by a factor of $\sim 15$.

\section{Experimental setup}
\label{sec:expsetup}

The data were collected with \kloe\ detector at \DAF, the Frascati \ff. 
\DAF\ is an \epm\ collider that operates at a 
center-of-mass energy of $\sim 1020\MeV$, the mass of the \Pphi\ meson.
\Pphi\ mesons decay $\sim 34\%$ of the time into nearly collinear \koko\ pairs. 
Because $J^{PC}(\phi)=1^{--}$, the kaon pair is in an antisymmetric state, 
so that the final state is always \ks\kl. 
Therefore, the detection of a \kl\ signals the presence of a \ks\ of known
momentum and direction, independently of its decay mode. 
This technique is called \ks\ {\it tagging}. A total of $\sim 4$ 
billion \Pphi\ were produced, yielding $\sim 1.4$ billion of \ks\kl\ 
pairs. 

The \kloe detector consists of a 
large cylindrical drift chamber (DC), surrounded by a lead/scintillating-fiber
sampling calorimeter (EMC). A superconductig coil surrounding the calorimeter
provides a $0.52\T$ magnetic field. The drift chamber \cite{drift}, $4\m$
in diameter and $3.3\m$ long, is made of carbon-fibers/epoxy and filled with a light gas mixture,
$90\%$ He-$10\%$iC$_4$H$_{10}$.
The DC position resolutions are 
$\sigma_{xy} \approx 150\um$ and $\sigma_{z} \approx 2\mm$.
DC momentum resolution is $\sigma(p_\perp)/p_\perp \approx 0.4\%$. Vertices
are reconstructed with a spatial resolution of $\sim3\mm$.

The calorimeter \cite{calo} is divided into a barrel and two endcaps
and covers $98\%$ of the solid angle.
The energy and time resolutions are $\sigma_E/E = 5.7\%/ \sqrt{E(\GeV)}$
and $\sigma_t=57\ps / \sqrt{E(\GeV)}\oplus 100\ps$, respectively.

To study the background rejection, a MC sample of \Pphi\ decays to all possible 
final states has been used, for an integrated luminosity of $\sim 1.9\Lfb$.
A MC sample of $\sim 45000$ signal events has been also produced, to measure the analysis efficiency.

\section{Data analysis}
\label{sec:ana}

The identification of \kl-interaction in the EMC is used to tag the presence
of \ks\ mesons. The mean decay lenghts of \ks\ and \kl\ are 
$\lambda_S \sim 0.6\cm$ and $\lambda_L \sim 350\cm$, respectively. About $50\%$
of \kl's therefore reach the calorimeter before decaying. The \kl\ interaction 
in the calorimeter barrel (\kcr) is identified by requiring a cluster of energy greater than $125 \MeV$
not associated with any track, and whose time 
corresponds to a velocity $\beta = r_{cl}/ct_{cl}$ compatible with the kaon
velocity in the $\phi$ center of mass, $\beta^* \sim 0.216$, after the 
residual $\phi$ motion is considered. Cutting at $0.17 \le \beta^* \le 0.28$ we selected 
$\sim 650$ million \ks-tagged events (\kcr\ events in the following), which are used as a starting sample for the 
\DKSee\ search. 
\\
 
\begin{figure}[t]
  \begin{center}
    \epsfig{figure=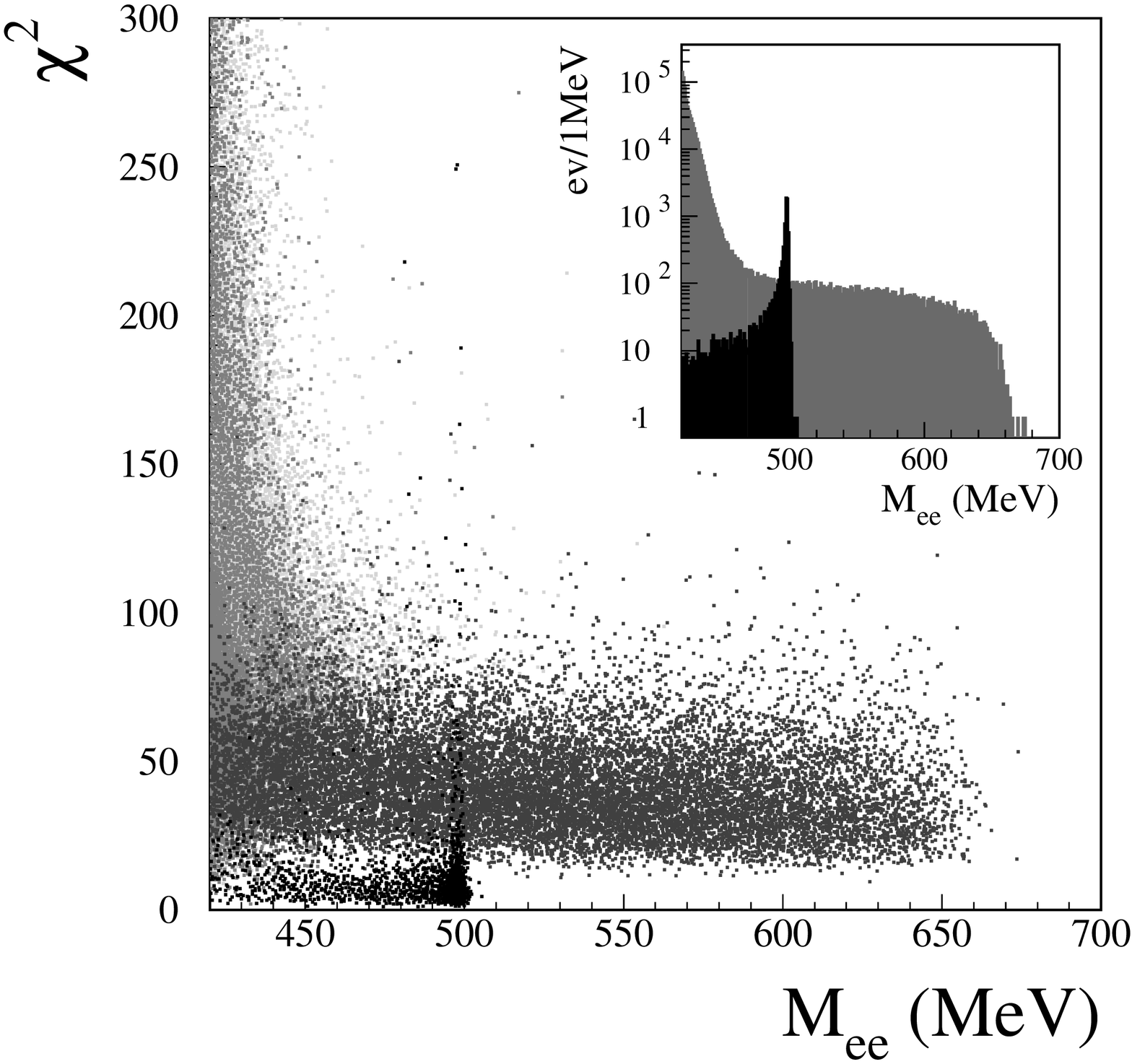,height=2.7in}
    \epsfig{figure=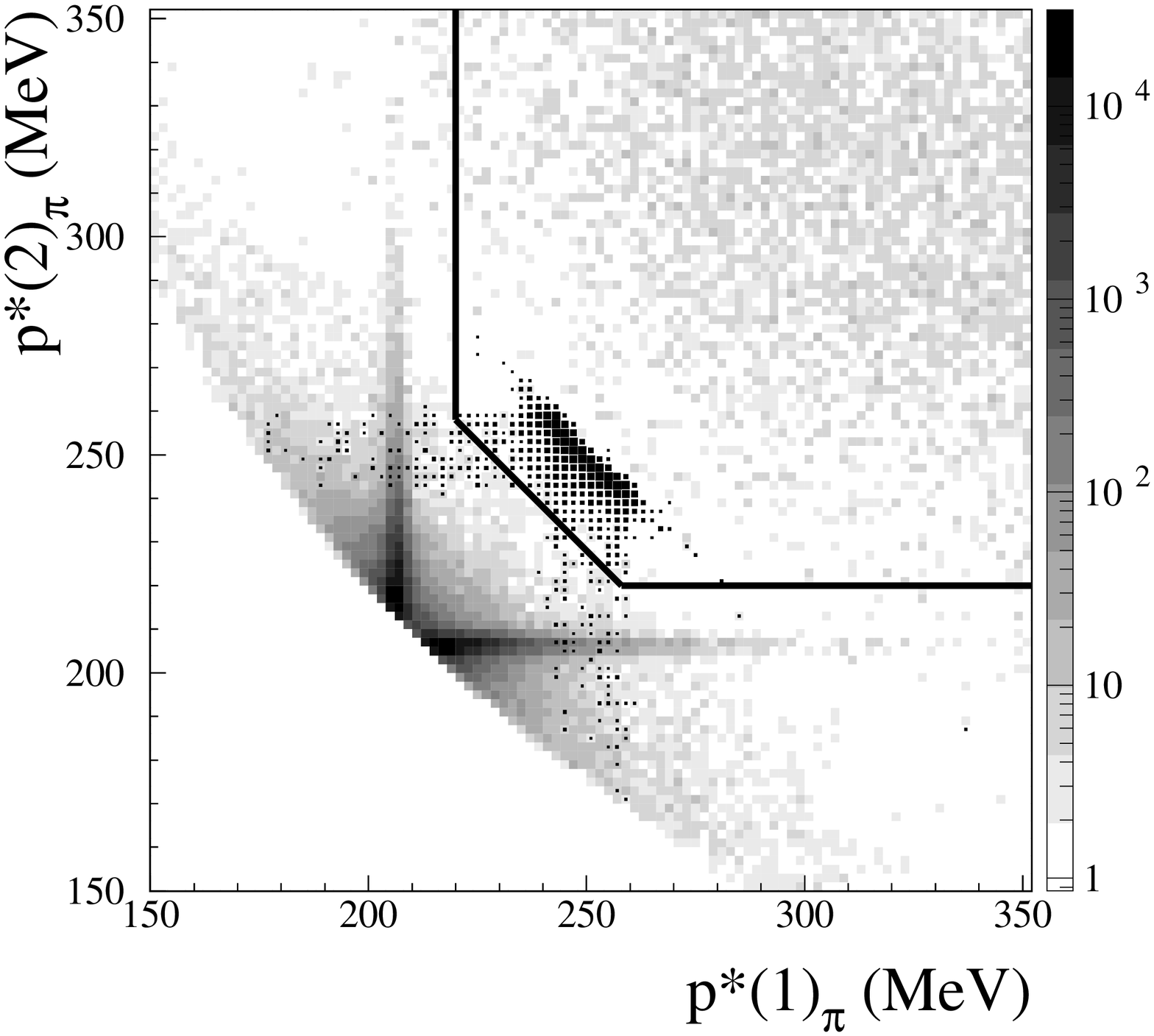,height=2.7in}

    \epsfig{figure=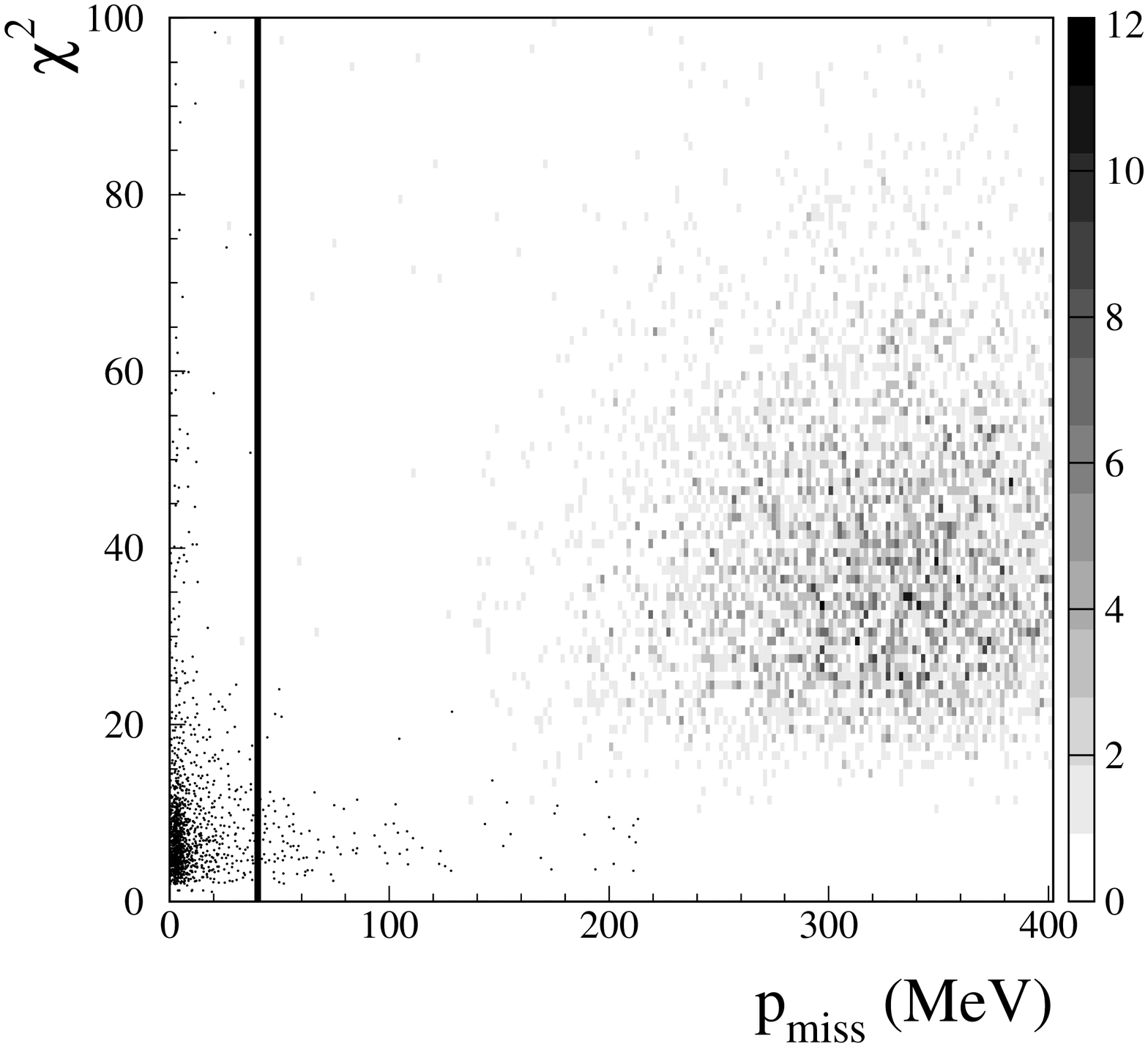,height=2.7in}
    \epsfig{figure=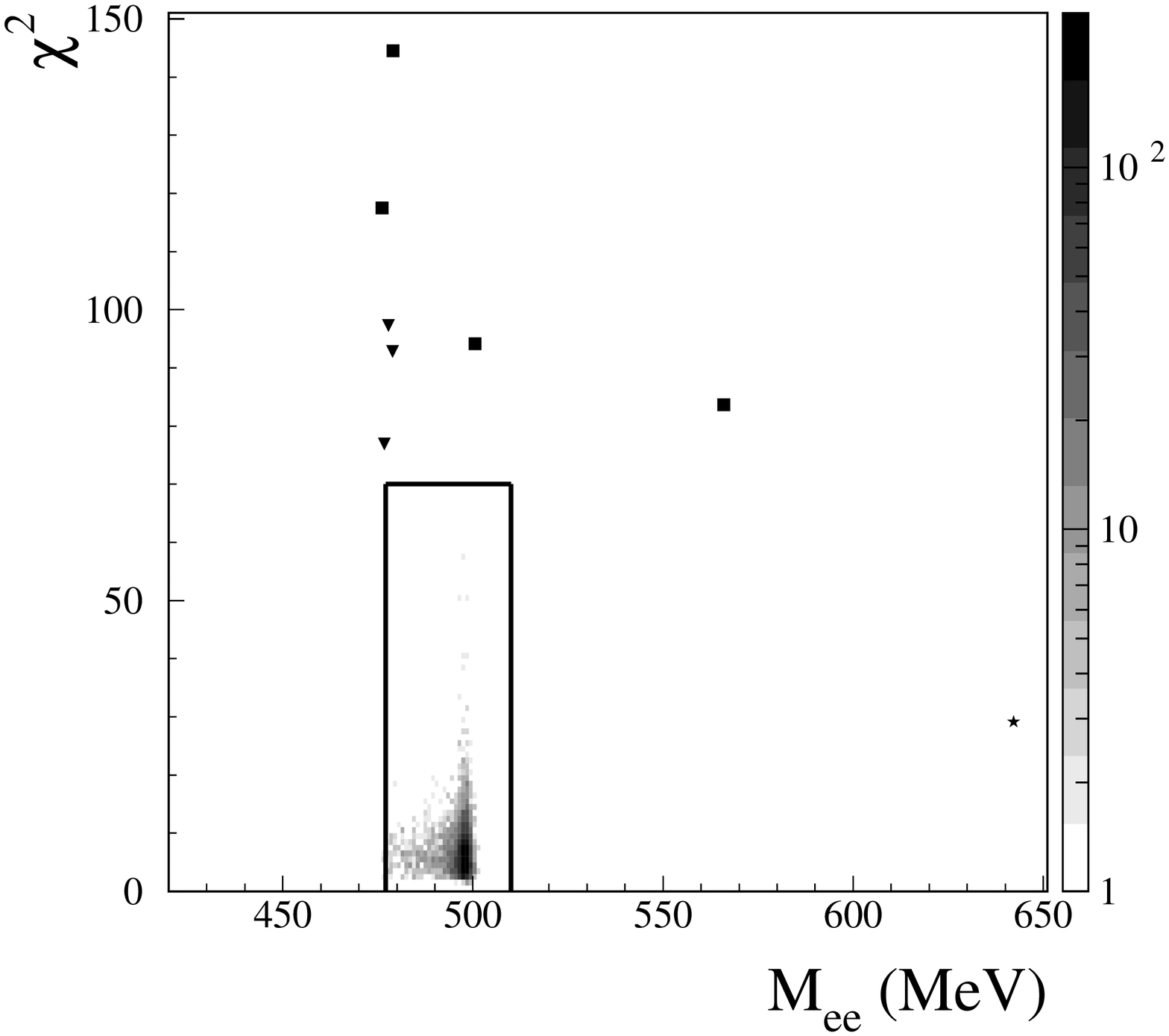,height=2.7in}
    \caption{Top left: $\chi^2$ vs $M_{ee}$ distributions for MC signal (black), MC backgrounds 
             \DKSpippim\ and \Dphipippimpio\ (light and dark grey respectively), 
             $M_{ee}$ distributions for MC signal (black) and MC backgrounds (grey) is shown in the inset;
             top right: $p_\pi^{\ast}$ distributions for MC signal (black) and MC background (grey scale);
             bottom left: $\chi^2$ vs $P_{\mathrm{miss}}$ distributions for MC signal (black) and 
             MC background (grey scale);
             bottom right: $\chi^2$ vs $M_{ee}$ distributions for MC signal (grey scale), data ($\blacksquare$),
             \DKSpippim\ ($\blacktriangledown$) and \Dphipippimpio\ ($\bigstar$) after background rejection cuts.}
    \label{fig:all}
  \end{center}
\end{figure}

\DKSee\ events are selected by requiring the presence of two tracks of opposite 
charge with their point of closest approach to the origin inside a cylinder $4\cm$ 
in radius and $10\cm$ in length along the beam line. 
The track momenta and polar angles must satisfy the fiducial cuts 
$120\le p \le 350\MeV$ and $30^{\circ} \le \theta \le 150^{\circ}$. 
The tracks must also reach the EMC without spiralling, and have an associated cluster.  
In Fig.~\ref{fig:all}, the two-track invariant mass evaluated in electron 
hypothesis ($M_{ee}$) is shown for both MC signal and background samples. 
A preselection cut requiring $M_{ee}> 420 \MeV$ has been applied, which rejects most
of $\DKSpippim$ events, for which $M_{ee}\sim 409 \MeV$. The residual background has 
two main components: \DKSpippim\ events, populating the low $M_{ee}$ region, and 
\Dphipippimpio\ events, spreading over the whole spectrum. The \DKSpippim\ events
have such a wrong reconstructed $M_{ee}$ because of track resolution or one pion decaying 
into a muon. The \Dphipippimpio\ events enter the preselection because of a machine 
background cluster, accidentally satisfying the \kcr\ algorithm. 
After preselection we are left with $\sim 5\times 10^5$ events.
To have a better separation between signal and background, a $\chi^2$-like 
variable is defined, collecting informations from the clusters associated
to the candidate electron tracks.
Using the MC signal events we built likelihood functions based on:
the sum and the difference of $\delta t$ for the two tracks, where 
$\delta t = t_{cl} - L/\beta c$ is evaluated in electron hypothesis;
the ratio $E/p$ between the cluster energy and the track momentum, for both charges;  
the  cluster depth, evaluated respect to the track, for both charges. 
In Fig.~\ref{fig:all}, the scatter plot of $\chi^2$ versus $M_{ee}$ is shown, for MC 
signal and background sources.
The $\chi^2$ spectrum for background is concentrated 
at higher values respect to signal, since both 
\DKSpippim\ and \Dphipippimpio\ events have pions in the final state.
\\

A signal box to select the \DKSee\ events can be conveniently defined in the 
$M_{ee}-\chi^2$ plane (see Fig.~\ref{fig:all}); nevertheless we investigated 
some more independent requirements in order to reduce the background 
contamination as much as possible before applying the $M_{ee}-\chi^2$ selection. 

Charged pions from \DKSpippim\ decay have a momentum in the \ks\ rest frame 
$p_\pi^{\ast}\sim 206\MeV $. The distribution of track momenta in the \ks\ rest frame, 
evaluated in the pion mass hypothesis, is shown in Fig.~\ref{fig:all}, for MC 
background and MC signal.  
For most of \DKSpippim\ decays, at least one pion has well reconstructed momentum, so that  
the requirements
\begin{equation} 
\mbox{min}(p^{\ast}_\pi(1), p^{\ast}_\pi(2))  \ge 220 \MeV \quad , \quad
p^{\ast}_\pi(1)+ p^{\ast}_\pi(2) \ge 478 \MeV
\end{equation}   
rejects $\sim 99.9\%$ of these events, while retaining $\sim 92\%$ of the signal.  

To reject \Dphipippimpio\ events we have applied a cut on the missing momentum,
defined as:
\begin{equation}
P_{\mathrm{miss}} = \left \vert \vec{P}_\phi - \vec{P}_L - \vec{P}_S \right \vert
\end{equation}
where $\vec{P}_{L,S}$ are the neutral kaon momenta, and $\vec{P}_\phi$ 
is the $\phi$ momentum. The distribution of $P_{\mathrm{miss}}$ is shown in Fig.~\ref{fig:all}, 
for MC background and for MC signal events. 
We require
\begin{equation}
P_{\mathrm{miss}}\le 40 \MeV\, ,
\end{equation}
which rejects almost completely the $3\pi$ background source which is 
distributed at high missing momentum.
\\

A signal box is defined in the $M_{ee}-\chi^2$ plane as shown Fig.~\ref{fig:all}. 
The $\chi^2$ cut for the signal box definition has been chosen 
to remove all MC background events: $\chi^2 < 70$.
The cut on $M_{ee}$ is practically set by the $p^*_\pi$ cut,
which rules out all signal events with a radiated photon 
with energy greter than $20\MeV$, 
correspondig to an invariant mass window:
$477 <  M_{ee} \le 510 \MeV$.
The signal box selection on data gives $N_{obs}=0$. 
The upper limit at $90\%$ CL on the expected number of signal
events is $UL(\mu_S) = 2.3$.

\section{Results}
\label{sec:re}

The total selection efficiency on \DKSee\ events is evaluated by MC, using the following parametrization:
\begin{equation}
\epsilon_{sig} = \epsilon(K_{crash}) \times \epsilon(sele \vert K_{crash})\, , 
\end{equation}
where $ \epsilon(K_{crash})$ is the tagging efficiency, and 
$ \epsilon(sele \vert K_{crash})$ is the signal selection
efficiency on the sample of tagged events. The efficiency evaluation includes 
contribution from radiative corrections. 
The number of \DKSpippim\ events $N_{\pi^+\pi^-}$ counted on the same sample of \ks\ tagged events
is used as normalization, with a similar expression for the efficiency. 
The upper limit on BR(\DKSee) is evaluated as follows:
\begin{equation}
 UL(BR(\DKSee)) = 
 UL(\mu_s)     \times \mathcal{R}_{tag}\times 
\frac{\epsilon_{\pi^+\pi^-}(sele \vert K_{crash})}{
\epsilon_{sig}(sele \vert K_{crash})}\times
\frac{BR(\DKSpippim)}{N_{\pi^+\pi^-}}, \nonumber 
\end{equation}
where $\mathcal{R}_{tag}$~\cite{kspipi} is the tagging efficiency ratio, corresponding to
a small correction due to the \kcr\ algorithm dependence on \ks\ decay mode,
and it is equal to $0.9634(1)$.
Using $\epsilon_{sig}(sele \vert K_{crash}) = 0.465(4)$,
$\epsilon_{\pi^+\pi^-}(sele \vert K_{crash}) = 0.6102(5)$ and 
$N_{\pi^+\pi^-} = 217,422,768$, we obtain
\begin{equation}
UL(BR(\DKSee(\gamma))) =  9.3 \times 10^{-9}, \; {\rm at} \;90\%\,{\rm CL}\, .
\end{equation}
Our measurement improves by a factor of $\sim 15$ on the
CPLEAR result~\cite{cplearksee}, for the first time including radiative
corrections in the evaluation of the upper limit.

\end{document}